\documentclass[%
a4paper,
onecolumn,
flushbottom,
12pt
]{article}
\pdfoutput=1
\usepackage{physics}
\usepackage{times}
\usepackage{amsmath}
\usepackage{amssymb}
\usepackage{dcolumn}%
\usepackage{bm}%
\usepackage{xspace}%
\usepackage{fancyhdr}   
\usepackage[inline]{enumitem}
\usepackage[symbol*,perpage,bottom,flushmargin,splitrule,multiple]{footmisc}
\DefineFNsymbols*{lamportnostar}[math]{\dagger\ddagger\S\P\|*{\dagger\dagger}{\ddagger\ddagger}}
\setfnsymbol{lamportnostar}
\usepackage[colorlinks=false]{hyperref}%
\usepackage[capitalise]{cleveref}%
\creflabelformat{equation}{#2#1#3} %
\usepackage[%
vmargin=2.0cm,
hmargin=1.5cm,
]{geometry}

\newcommand{\expertBC}{{\itshape BC}\xspace}
\newcommand{\expertDK}{{\itshape DK}\xspace}
\newcommand{\expertDvD}{{\itshape DvD}\xspace}
\newcommand{\expertEW}{{\itshape EW}\xspace}
\newcommand{\expertLL}{{\itshape LL}\xspace}
\newcommand{\expertMB}{{\itshape MB}\xspace}
\newcommand{\expertSA}{{\itshape SA}\xspace}
\newcommand{\expertSI}{{\itshape SI}\xspace}

\newcommand{\phystatnu}{{PhyStat-$\nu$}\xspace}
\newcommand{\phystatnuipmu}{{PhyStat-$\nu$ IPMU}\xspace}

\usepackage{enumitem}
\usepackage{float} %

\usepackage[resetlabels]{multibib} %
\newcites{PSnu}{\phystatnuipmu Presentations}
\bibliographystylePSnu{phystatnu}

\usepackage{authblk} %

\newcommand{\nova}{\ensuremath{\textrm{NOv\hspace{-0.05em}A}}\xspace}
\newcommand{\novabold}{\ensuremath{\textrm{\bfseries NOv\hspace{-0.05em}A}}\xspace}
\newcommand{\nLEM}{\ensuremath{N_\textrm{LEM}}\xspace}
\newcommand{\nLID}{\ensuremath{N_\textrm{LID}}\xspace}
\newcommand{\nLIDiLEM}{\ensuremath{N_\textrm{LID$\cap$LEM}}\xspace}
\newcommand{\NH}{\ensuremath{\mathrm{NH}}}
\newcommand{\IH}{\ensuremath{\mathrm{IH}}}

\setlist[1]{itemsep=-0.2em} 
\setlist[2]{itemsep=-0.2em} 
\setlist[3]{itemsep=-0.2em} 
\setlist[4]{itemsep=-0.2em} 
\setlist{parsep=-0.5em} 
\setlist{partopsep=-0.5em} 
\setlist{topsep=-0.0em} 

\makeatletter
\g@addto@macro\bfseries{\boldmath}
\makeatother

\begin{document}
\flushbottom

\title{\phystatnu 2016 at the IPMU: Summary of Discussions}%

\renewcommand\Authfont{\normalsize}
\renewcommand\Affilfont{\normalsize}

\author[1]{Yoshi Uchida}
\author[2]{Mark Hartz}
\author[1]{R. Phillip Litchfield}
\author[3]{Callum Wilkinson}
\author[4]{Asher Kaboth}
\affil[1]{Imperial College London, United Kingdom}
\affil[2]{Kavli IPMU, Kashiwa, Japan}
\affil[3]{University of Bern, Switzerland}
\affil[4]{Royal Holloway University of London, United Kingdom}

\date{\today}%

\maketitle
\begin{abstract}
The presentations, discussions and findings from the inaugural ``\phystatnu'' workshop held at the Kavli Institute for the Physics and Mathematics of the Universe (IPMU) near Tokyo in 2016 are described.
\phystatnu was the first workshop to focus solely on statistical issues across the broad range of modern neutrino physics, bringing together physicists who are active in the analysis of neutrino data with experts in statistics to explore statistical issues in the field.
It is a goal of \phystatnu to help serve the neutrino physics community by providing a forum within which such statistical issues can be discussed and disseminated broadly.

This paper is adapted from a summary document that was initially circulated amongst the participants soon after the workshop.
Another \phystatnu workshop is being held at CERN in January 2019, building on the discussions in 2016.

Advances in experimental neutrino physics in recent years have led to much larger datasets and more diversity in the properties of neutrinos that are being investigated.
The discussions here raised several areas where improved statistical errors and more complicated interpretations of the data require statistical methods to be revisited, as well as topics where broader discussions between experimentalists, phenomenologists and theorists will required, which are summarised here.
It is important to record the state of the field as it stands today, as much is expected to change over the coming years, including the emergence of more inter-collaborational studies and increasing sophistication in global parameter fitting and model selection methods.
The document is also intended to serve as a reference for pedagogical material for those who are new to the use of modern statistical techniques to describe experimental data, as well as those who are well-versed in these techniques and wish to apply them to new data. 
\end{abstract}

\tableofcontents

\section{\label{sec:Introduction}Introduction}
The inaugural \phystatnu workshop was held at the Kavli IPMU over three days from the 30th of May, 2016.
Ninety registered participants from 14 countries engaged in talks and discussion, with 21 plenary presentations in addition to several introductory, keynote and summary talks~\citePSnu{Hartz,Uchida,Petcov,VanDyk,Cousins}\footnote{References to \phystatnuipmu presentations are given in this style in this document.}, as well as short talks from the contributed poster presenters.
On the final day, a panel discussion was held to discuss some open issues and questions submitted by the participants.

The bulk of the participants were active neutrino physicists, but six of those present were designated as ``invited experts''---some in pure statistics as well as those in the statistical treatment of particle physics data through their roles in past ``collider'' PhyStats---to help inform the discussions and provide their own insights on the methods being used and proposed by the neutrino physicists. 

The invited experts were the statisticians Sara Algeri (Imperial College London), Michael Betancourt (University of Warwick), David Van Dyk (Imperial) and Shiro Ikeda (Institute for Statistical Mathematics), and the physicists Bob Cousins (UCLA) and Louis Lyons (Imperial)\footnote{The invited experts and panel members are henceforth referred to by their initials in italics.}. 
The panel members for the discussion in the closing session were \expertSA,
\expertMB and Dean Karlen (University of Victoria) and Elizabeth Worcester
(Brookhaven National Laboratory), with Yoshi Uchida (Imperial) as chair.

In this paper, we briefly summarise the talks, discussions and findings, and list the open issues that would benefit from further discussion in the community\footnote{The website for \phystatnuipmu, at \url{http://conference.ipmu.jp/PhyStat-nu}, contains the full record of the workshop, including the presentation files.}. 
The format of the workshop and order of the talks is not followed here, but rather, contributions from the presentations, discussions, panel session and summary talks are grouped together by topic.
We also include selected references to reading material, both for introductory purposes to help those who are unfamiliar with the statistical methods discussed here, and as further reading on the more advanced topics which are referred to.

It is the aspiration of the organisers that this and future workshops in the series\footnote{A second \phystatnu was held at Fermilab from the 19th of September 2016.} and other modes of \phystatnu discussion will continue to provide a forum to help the neutrino physics community make the most of modern statistical methods as its experiments and theories make significant advances in the near future.

\section{\label{sec:StatisticsOverview}Overview of Statistical Concepts}
The workshop was preceded by an introductory lecture on statistical issues~\citePSnu{Lyons}, whilst prior reading material on statistics had also been recommended to participants, as given in \cref{sec:RecommendedReading}\footnote{A list of recommended reading for statisticians, on the basic concepts needed to acquaint oneself with the treatment of neutrino data was also provided, which can be found at \href{http://conference.ipmu.jp/PhyStat-nu}{the workshop website}}.
In this section we aim to provide some indication of the concepts that need to be appreciated to participate in discussions at the level of those at \phystatnu.

\paragraph{Essential Statistics}
The topics that were found to be most relevant to the discussions that occurred at the workshop include (page numbers are given for the relevant section of the introductory lecture~\citePSnu{Lyons}):
\begin{itemize}
  \item Likelihood Distributions (page~4)
  \item Coverage (page~9)
  \item Neyman Construction (page~23)
  \item Feldman-Cousins (Unified Approach) Construction (page~27)
  \item Hypothesis Testing (page~31)
	\begin{itemize}
	  \item null hypotheses (page~32)
	  \item Type-I and Type-II errors\footnote{Type I error: a ``false positive'', where a true null hypothesis is incorrectly rejected; Type II error:  conversely, a ``false negative'', where a false null hypothesis is incorrectly accepted.}
	  \item $p$-values and statistical power (page~35)
      \item Nested and Non-Nested Hypotheses (page~49)
      \item Simple vs Simple testing\footnote{When each of the two hypotheses being tested are ``simple'', i.e., are single distributions with no free parameters.}
	\end{itemize}
  \item Wilks' Theorem (page 48)
  \item Bayesian Statistics and choices of ``priors'' (page~18) and coverage. 
  \item The Jeffreys-Lindley Paradox (page 62; see also~\cite{Cousins:2013hry})
\end{itemize}
A good understanding of these concepts will provide the foundations for discussions on the current state of the use of statistics in neutrino physics.

\section{\label{sec:talksanddiscussion}Discussions and Findings} %
We describe here the broad areas of neutrino physics and statistics that were
discussed at \phystatnuipmu. No attempt is made to introduce the details
of experimental neutrino physics in this document, for which we refer the 
reader elsewhere\footnote{See, for example, the
presentation files from the Neutrino 2016 conference 
for up-to-date reviews on the status of the field:
\url{http://neutrino2016.iopconfs.org/programme}.}.
References are chosen for their pedagogical qualities; original physics documents can be found by following the information in the presentation files.

  \subsection{\label{sec:LongBaseline}Long-Baseline Accelerator Experiments}

Long-baseline (LBL) accelerator neutrino oscillation experiments are pursuing the measurement of five quantities: $\theta_{23}$, $\theta_{13}$, $\Delta m^2_{32}$, $\mathrm{sign}(\Delta m^2_{32})$ and $\delta_\text{CP}$~\citePSnu{Kaboth}\footnote{The validity of the three-neutrino model upon which rests is also another important topic of study.}. The parameters $\theta_{23}$ and $\Delta m^2_{32}$ are most directly connected to $\nu_{\mu}\rightarrow\nu_{\tau}$ oscillation (most often observed as $\nu_\mu$ disappearance), and the other parameters are accessible via $\nu_{\mu}\rightarrow\nu_e$ oscillation. Atmospheric measurements are complementary in $\theta_{23}$ and $\Delta m^2_{32}$, and reactor measurements are complementary in $\theta_{13}$. These experiments are now pursuing $\delta_\text{CP}$, the unknown CP-violating parameter, and $\mathrm{sign}(\Delta m^2_{32})$, the unknown mass hierarchy (MH). The MH is a topic of significant broader interested (see \cref{sec:MassHierarchy}), and here we focus on the overall tools used for currently-running LBL experiments and especially CP violation. 

\subsubsection{\label{sec:LongBaselinet2k}T2K}

T2K is an experiment using the neutrino beam from the \mbox{J-PARC} accelerator laboratory, with a baseline of 295~km. In terms of statistical methods, T2K has taken an approach with a variety of analyses using both frequentist and Bayesian methods. T2K also has two analysis streams, one of which fits near detector data separately and passes the information on as a set of multi-dimensional Gaussian constraints, and one of which simultaneously fits near and far detector data. 

For the main $\delta_\text{CP}$ analyses, there are two methods that are used to assess the interval constraint on the parameters: a frequentist method which uses a Feldman-Cousins procedure~\cite{Feldman:1997qc} to determine critical values as a function of $\delta_\text{CP}$, where it is noted that the critical values found are different from the asymptotic case by around 0.5 units of $\Delta\chi^2$; and a Bayesian method that uses a Markov Chain Monte Carlo (MCMC) using the Metropolis-Hastings algorithm and produces a highest posterior density interval in $\delta_\text{CP}$. Although the exact boundaries using these two methods are slightly different, the broad physics conclusions from them are in agreement. There was a consensus among the workshop participants that the use of many different tools~\citePSnu{Sgalaberna, Shah, Haegel} was a benefit to the analyses.

\paragraph{Elimination of nuisance parameters \label{para:LBLmarginalisationprojection}}The workshop considered a number of special issues related to T2K. The first of these was that there had been a long internal discussion in T2K over whether to use marginalisation or profiling over systematic parameters, as this had been seen to be a significant effect in some analyses. Over time, T2K has moved to marginalisation over systematic parameters, on the understanding that this is a more correct representation of uncertainties, especially in cases where there are non-Gaussian distributions.  The discussion also made reference, alongside marginalisation and profiling, to what was termed ``projection''---meaning the use of a single fixed value for the nuisance parameter.  With regard to projection, \expertMB noted during the discussion: 
\begin{quote}
Let $p(x, y)$ be a joint density: the question was how to convert this into a density over $p(x)$ alone. ``Marginalisation'' is defined as
\begin{equation}
  p_\textrm{m} (x) = \int p(x, y) \dd{y} = \int p(x | y) p(y) \dd{y},
\end{equation}
whilst ``projection'' is defined as 
\begin{equation}
  p_\text{p} (x) = p(x, y_\textrm{max})  =  \int p(x | y) \delta( y - y_\textrm{max}) \dd{y}.
\end{equation}
For any $p(y)$, $p_\text{m} (x)$ will always have a larger variance than $p_\text{p} (x)$ and hence is more conservative. The marginal also happens to be the correct probabilistic way to incorporate uncertainty about $y$ into $x$ from a purely mathematical perspective.
\end{quote}
The issue of how to eliminate nuisance parameters is a long-standing problem in statistics and was revisited within other contexts (\cref{subsec:overallprofilingmarginalisation})---and the importance of emphasising which method is being used in any particular analysis was highlighted throughout the discussions.

\paragraph{Prior- and posterior-predictive distributions\label{para:priorpredictive}}The second special issue was about constructing a $p$-value in the presence of high dimensionality systematic uncertainty, especially concerning the case of testing the hypotheses of no CP-violation versus CP-violation. Discussion centred on whether toy MC experiments should be thrown with the prior distributions of systematic parameters (so-called prior-predictive) or post-fit distributions (so-called posterior-predictive)\footnote{
The posterior predictive distribution is defined as $p(y_\textrm{new} | y) = \int p(y_\textrm{new} | \theta)
p(\theta | y) \dd{\theta}$, where $p(y_\textrm{new} | \theta)$ is the likelihood for a new measurement and $p(\theta | y)$ is the posterior.}, in particular concerning other oscillation parameters that may be constrained in the fit.  Both were considered to have significant problems: the prior-predictive method speaks to the whole model, not just the parameter of interest, and so can be skewed by a bad prior choice. However, the posterior-predictive method is not strictly a significance test, as there is no way to calculate the power of the test. The general conclusion was that there is no easy way to build a $p$-value in the way that is desired. 

\paragraph{Bayes factors\label{para:bayesfactors}}Later in the workshop there was discussion of how to create a Bayes factor for these hypotheses; the general consensus was that the intervals are probably sufficient in this case---exclusion at a high probability level excludes those parameter values sufficiently. A Bayes factor can be constructed with the Savage-Dickey ratio~\cite{DickeyLientz:DickeySavageRatio}%
, but this is less preferred. 

The final special issue for T2K was the use of alternate cross section models in oscillation analyses, which is discussed in detail in \cref{sec:CrossSections}.

\subsubsection{\label{sec:LongBaselinenova}\novabold}

\nova is an experiment using the NuMI beamline at Fermilab, and a baseline of 810~km~\cite{Ayres:2007tu}. At the time of the workshop, \nova had presented separate $\nu_{\mu}\rightarrow\nu_{\mu}$ and $\nu_{\mu}\rightarrow\nu_{e}$ analyses, but not a combined analysis, unlike T2K. Both \nova analyses use an unfolding and re-folding of near detector data
to constrain flux and cross section information, and both %
analyses use frequentist methods to draw confidence intervals and regions, using Feldman-Cousins in the $\delta_\text{CP}$ variable for the $\nu_{\mu}\rightarrow\nu_{e}$ analysis.

The most significant discussion about \nova came from its use of two different event selections for the $\nu_e$ analysis, and the fact that one selection (library event matching, or "LEM") resulted in $\nLEM=11$ signal events, and the other (likelihood discriminator, or "LID") found $\nLID = 6$ events, despite the Monte Carlo predicting approximately the same number of events for both selections. \nova devised a procedure to determine if this result was an unreasonable statistical excursion of the data, by calculating all of the combinations of $n=(\nLID-\nLIDiLEM)$, $m =(\nLEM-\nLIDiLEM)$, and $I = \nLIDiLEM$ for a given number of background events, $B$, with the trinomial probability:
\begin{equation}
P_B(n, m, I) = \frac{N!}{n!\,m!\,I!}\, p^{\textrm{LID}}_B(n) p^{\textrm{LEM}}_B(m) p^{\textrm{LID$\cap$LEM}}_B(I).
\end{equation}
The data are $n=0$; $m=5$; $I=6$ and the $P_B$ for all 
combinations are ordered by probability and the sum of all the probability for cases less probable than the data is calculated. This is repeated for $B=0,1,2,\ldots$ and then a weighted sum over $B$, where the probability of the number of background events is calculated from Monte Carlo, is found. By carrying out this procedure, \nova finds that the probability of a more extreme event is 7.8\%, indicating that there is some tension, but that it is not too extreme\footnote{Reports from \nova at Neutrino 2016 indicated that these two selections have been re-optimised for the somewhat larger data sample, and, after this tuning, are now in closer agreement. Moreover, \nova analyses have since moved to a different event selection algorithm altogether.}.  %

\paragraph{Future experiments}While the focus at this workshop was on the current generation of long-baseline experiments, there is much ongoing work on designing future experiments---an example is the work on NuPRISM~\citePSnu{Yoshida}.

  \subsection{\label{sec:CrossSections}Neutrino-Nucleus Interaction Cross Sections}
\paragraph{The use of cross sections in oscillation experiments}Neutrino oscillation experiments measure the event rate $R(\vec{\boldsymbol{x}}')$ as a function of reconstructed particle kinematics, $\vec{\boldsymbol{x}}'$, and infer the  oscillation probability $P(\nu_{A} \rightarrow \nu_{B})$, which depends on the neutrino energy, $E_{\nu}$:
\begin{equation}
  R(\vec{\boldsymbol{x}}') = \int \dd{E_\nu} \Phi(E_\nu) \cdot \sigma(E_{\nu}, \vec{\boldsymbol{x}}) \cdot \epsilon(\vec{\boldsymbol{x}}, \vec{\boldsymbol{x}}')\cdot P(\nu_{A} \rightarrow \nu_{B}; E_\nu),
  \label{eq:rate}
\end{equation}
\noindent which relies on an understanding of the flux prediction $\Phi(E_\nu)$, the cross section $\sigma(E_{\nu}, \vec{\boldsymbol{x}})$
(which relates $E_{\nu}$ with the true outgoing particle kinematics, $\vec{\boldsymbol{x}}$)
and the detector response (smearing) $\epsilon(\vec{\boldsymbol{x}}, \vec{\boldsymbol{x}}')$.

Accelerator-based long baseline oscillation experiments~\citePSnu{Kaboth} rely on flux, cross section and detector models, which have prior uncertainties set by external measurements. Typically a near detector is used to constrain these models with an unoscillated event rate, and a far detector, where the oscillated event rate is used to extract oscillation parameters by marginalizing over the entire likelihood given by the flux, cross section and detector models. For atmospheric neutrino experiments~\citePSnu{Bruijn, deAndre}, there is no near detector to constrain the input models.

Interaction-level quantities (e.g. $E_{\nu}$, $Q^{2}$, energy transfer, etc.) cannot be reconstructed, so model-independent cross section measurements are only possible in terms of differential final state particle kinematics. In order to constrain cross section models which care about interaction-level quantities, it is therefore necessary to combine data from many experiments. Unfortunately, the available cross section models used by current experiments cannot consistently describe the global dataset~\citePSnu{Wascko}, which presents a problem for oscillation experiments, and ultimately leads to model-dependent (and possibly biased) oscillation results. A comment was made (\expertMB) that often, a larger model can be found for which the available models represent particular values of this larger model. Unfortunately, the cross section models available work from very different fundamental assumptions, so it is difficult to see how a larger and physically meaningful model could be found that would satisfy this need. Finding a theoretical model which describes neutrino cross sections well, with parameters well-constrained by experiment, is essential for precision neutrino oscillation measurements and their statistical interpretation. Another suggestion from \expertLL was a method called ``discrete profiling'' that was used at the LHC in the context of handling uncertainties in background shapes~\cite{1748-0221-10-04-P04015}.

\paragraph{Modelling neutrino interactions}The simplest interaction that is relevant for the few-GeV energies of interest to most long-baseline oscillation experiments is the Charged-Current Quasi-Elastic (CCQE) interaction ($\nu_{l} + n \rightarrow l^{-} + p$), for which there have been a number of recent measurements. Efforts to select a cross section model from those available and to constrain the parameters of the chosen model using the global CCQE dataset were described~\citePSnu{Wilkinson} (see also~\cite{Wilkinson:2016wmz}). A number of difficulties must be overcome, largely due to problems with the way available data are presented. As is typical for modern neutrino cross section measurements, backgrounds to the signal are subtracted by each experiment---using their own background model---and the results are unfolded to the true particle kinematics to remove detector smearing, often using the D'Agostini method~\cite{D'Agostini:1994zf}. For some datasets, correlations between histogram bins are not available. Many similar issues to this have been encountered before by parton distribution function fitting groups (see, for example,~\cite{pumplin2000-pdfs}), and the analysis presented here adopts many of the same strategies (such as taking data at face value and using an ad-hoc inflation of uncertainties).%

\paragraph{Bayesian hierarchical modelling\label{para:bayesianhierarchicalmodelling}}A radically different approach, suggested by \expertMB and \expertDvD, is to perform a meta-analysis using Bayesian hierarchical modelling. The basic principle is to construct a model for the cross section where the data are the results from each experiment. The published central values and uncertainties (diagonals of the covariance matrix) are treated as data and inform the global fit. An example of a meta-analysis using a similar technique can be found in~\cite{DvD-hierarchical-bayes}.
However, the fact that the data are unfolded presents a further complication to the use of techniques such as this, and although this method should give statistically sound uncertainties without resorting to drastic measures such as ad-hoc uncertainty inflation, it was not clear to the experts how to select one model from a range using such methods. A challenge for this technique would be the subjective choices required to construct a data model; in cases with only the diagonals of the correlation matrix to hand, very little information is available, so the initial choices may strongly affect the outcome. However these choices still are more rigorous than those used in the naive uncertainty inflation analyses.  It was also noted by \expertBC and \expertLL that analyses of this nature are very challenging and would probably require an expert for each detector; pragmatically, the current approach, similar to the parton density fitting groups, may be the best that can be done with the current dataset. In contrast to this, \expertMB noted that these analyses can in fact be quite straightforward, especially in collaboration with a statistician who performs similar analyses in other areas of study as a matter of routine.

A related issue for fitting models to existing data was presented~\citePSnu{Stowell} wherein some cross section analyses give many different results (differential cross sections with respect to different kinematic variables), but often do not provide correlations. In this case, the set of distributions used in a fit will affect the results. This is taken care of naturally by the data model in the meta-analysis approach, but for the less sophisticated methods currently used, the advice was to simply fit all distributions, and then assess the result. %
As the distributions are generally all unfolded separately, trying to guess the correlations after the fact is non-trivial.

\paragraph{Cross section measurements}An overview of neutrino cross section measurements was presented~\citePSnu{Wascko}. The ``traditional'' way to calculate a flux-averaged differential cross section in the $i^\textrm{th}$ bin of a true kinematic variable $x$, $\dv*{\sigma}{x_{i}}$ is
\begin{equation}
  \dv{\sigma}{x_{i}} = \frac{\sum_{j} \widetilde{U}^{-1}_{ij} \left(N_{j} - B_{j}\right)}{\Phi_{\nu}T\Delta x_{i} \epsilon_{i}}
  \label{eq:xsec}
\end{equation}
\noindent where $N_{j}$ ($B_{j}$) is the number of selected (predicted background) events in reconstructed bin $j$, $\Phi_{\nu}$ is the total integrated flux, $T$ is the number of target nuclei per unit area, $\Delta x_{i}$ is the width of the true bin, $\epsilon_{i}$ is the selection efficiency, and $\widetilde{U}^{-1}_{ij}$ is the unfolding matrix (the pseudoinverse of the matrix, $U_{ij}$ which describes detector smearing). This is a rearrangement of \cref{eq:rate}, where the rate is the number of selected events, the unfolding matrix accounts for the detector smearing, and $\epsilon_{i}$ accounts for detector and selection inefficiencies. The flux and other kinematic variables have been integrated over in \cref{eq:xsec}. Limiting the unsmearing matrix $\widetilde{U}^{-1}_{ij}$ and the efficiency $\epsilon_{i}$ to only correct for the detector Monte Carlo quantities---rather than also correcting for cross section processes---would reduce the model dependence of the results, and make them easier to use.

\paragraph{Publishing cross section results\label{para:crosssectionpublishedmeasurement}}The information published as a cross section measurement (using \cref{eq:xsec}) and available to theorists is much less complete than the information used by experiments themselves for oscillation analyses (\cref{eq:rate}). To bridge this divide, two alternative\footnote{But not without precedent; see, for e.g.,~\cite{AguilarArevalo:2010cx} and~\cite{AguilarArevalo:2008rc,AguilarArevalo:2010wv}} ideas for how to present data were proposed:
\begin{itemize}
  \item {\bfseries Idea 1} Publish uncorrected data, the smearing matrix $U_{ij}$ and efficiency functions to allow users of the data to smear their theory to match the data.
  \item {\bfseries Idea 2} Apply efficiency weights and smearing event-by-event and provide an ntuple of the data, with the normalisation constants $\Phi_{\nu}T$ included for correct normalisation.
\end{itemize}
Both ideas are strongly supported by all of the experts present, and much of the discussion during the conference pertinent to cross section measurements focused on unfolding, and avoiding data reduction, both of which relate to these ideas.
It should be noted that there needs to be discussion among all stakeholders (experimentalists who produce and use cross-section results, and theorists who predict and study them) regarding how to move forward, and alongside the statistics reasoning that may dominate at \phystatnu, the broader community of neutrino physicists should be expected to take part (e.g., through the NuSTEC~\cite{Morfin-NuSTEC-nuint15} group).

More specifically, \expertMB commented that the smearing and efficiency functions with MC samples---Idea 1 above---is exactly the same as publishing a likelihood, and that look-up tables are too inaccurate in higher dimensions.

\expertEW noted that for sensitivity calculations, DUNE are looking at publishing a more open toolkit, where one specifies a ``configuration'' to allow the set-ups, including the latest ``best'' one, to be used. This allows others to argue for, and insert, different assumptions. This also corresponds to Idea 1 above. She added that the phenomenology community are very excited by this. It was initially considered that this may be a difficult concept to implement, but there has been almost no resistance from the collaboration. It should be noted that this may be because at this point in time this only involves simulation studies, not real data.

\paragraph{Unfolding of detector effects\label{para:unfolding}}Unfolding, or deconvolution, is the procedure for removing from the data the smearing effects caused by imperfect detectors~\cite{Prosper:2011zz}. In the simplest case, one simply inverts the smearing matrix $U_{ij}$ to obtain $U^{-1}_{ij}$, but this often leads to large oscillations in the unfolded data as noise in the smeared signal is blown up in the inversion~\cite{kuuselamaster}. Various methods for smoothing (or regularizing) the resulting unsmeared distributions exist, where a pseudoinverse matrix $\widetilde{U}^{-1}_{ij}$ is calculated instead. The most common method used currently in neutrino physics is D'Agostini unfolding~\cite{D'Agostini:1994zf}, known as Bayesian unfolding\footnote{As indicated by the titles of the original papers.}. However, it was argued by \expertBC that the method is not Bayesian, or unique to D'Agostini, and is simply an iterative method for calculating the maximum likelihood estimate of the inverse, which would be equivalent to matrix inversion in the Gaussian limit when the method converges~\cite{kuuselamaster}---although in practise, regularisation of the solution is achieved by stopping after a small number of iterations. The uncertainty from the unfolding procedure is difficult to quantify accurately when the method is stopped after an arbitrary number of iterations. 

The unambiguous advice from the experts present was {\itshape not to unfold unless absolutely necessary,} and that it was not necessary to unfold in this case. The collider community is moving away from unfolding, and the general advice is to smear the theory to match the data, rather than try to unsmear the data~\cite{Cousins:2016ksu}. If one must unfold, various alternative methods do exist for smoothing the data which may be preferable to the D'Agostini method~\cite{kuuselamaster}. Idea 1 presents the data simply without the unfolding step, and would avoid the pathologies of unfolding, which can make downstream analyses more difficult (see for example~\cite{Wilkinson:2016wmz}).

\paragraph{Generative modelling\label{para:generativemodelling}}\expertMB presented a talk on generative modelling~\citePSnu{Betancourt2}, where it was argued that cross section experiments should aim to publish the full analysis model, or likelihood, composed of the conditional probability distributions calculated for each step in the analysis, rather than presenting only the mean and variance for each source of uncertainty. From a statistician's point of view, the advantage of this approach is that it removes a false distinction between relevant and nuisance parameters and treats all stages of the data generation process equally. The practical advantage for physicists is that future users of the data could change some elements of the model and not have to rely solely on the model used by the experiment at the time of publication, or to ensure that models are consistent between multiple experiments when performing meta-analyses. Whilst in an ideal case every step of the analysis would be presented in the form of probability distributions, in practice it is reasonable to start implementing this for single steps that occur in the analysis chain. For neutrino oscillation analyses, the full likelihood for \cref{eq:rate} is used, so this information is in principle already available to experiments, and Idea 2 is a proposal to release that information. 

\paragraph{Data release tools\label{para:datareleasetools}}Some concerns were raised about the practicality of releasing so much information, but dedicated software is available~\cite{stan-manual, Kruschke:2010:DBD:1951940}, and efforts to release similar levels of data are in progress at the LHC. The experts agree that any steps towards publishing the full likelihood, either as an approximate likelihood or incomplete parts of the likelihood, would be useful for downstream users of the data.

\subsection{\label{sec:ReactorNeutrinos}Reactor Neutrinos}
A review of the statistical methods being used at short- and medium-baseline reactor neutrino experiments was presented~\citePSnu{Seo}.
The individual $O(1)$~km-baseline experiments  employ $\chi^2$ fits for parameter estimation, with differing implementations: covariance matrix (Daya Bay), the pull method (RENO), and a hybrid method (Double Chooz).
The experiments are moving towards combining their results to produced uncertainties that are below 3\%, and advice was sought; while it is much easier to simply combine the $\chi^2$s, the recommendation was (\expertLL) to performed a full combined analysis\footnote{At the LEP collider, this required a year of weekly discussions.}.

Some discussion centred on the oscillation contour plots that are presented by the experiments.
Here too, clarification on whether marginalisation or profiling is used to produce one-dimensional distributions was requested of the experiments\label{para:reactorneutrinoprofiling}.
Another comment (\expertDvD) was that when contours do not look Gaussian (a common issue with oscillation parameter contours), it is inadvisable to show a 99.7\% region without a implementing a full MC study.

While the bulk of the reactor neutrino discussions centred on 1~km and 50~km experiments, the new generation of very short-baseline experiments was also represented in the form of the 24~m-baseline NEOS experiment~\citePSnu{Ko}.

\subsection{\label{sec:DirectNeutrinoMass}Direct Neutrino Mass Measurements}
Direct neutrino mass measurements, which take the form of studying the shape of the high-energy tail of a $\beta$-decay spectrum, are a quintessential example of the canonical bounded Gaussian problem~\cite{Cousins:BoundedGaussian}, with the squared neutrino mass being the most natural quantity to be measured~\citePSnu{Kleesiek}.

Here again, $\chi^2$ minimisation has been the most common approach, the ongoing efforts to introduce Bayesian methods in this area were also described.
The issues that this raised include the choice of prior (flat in $m_\nu$ or in $m^2_\nu$?) and the problem of the non-physical region\footnote{This resulted in the following exchange:
Physicist: ``it pains me to have to model the non-physical region''---Statistician: ``it pains me more that you do this''.}, where it was proposed that a prior that is 0 for $m^2_\mu < 0$ be used. The comment from \expertBC was:
\begin{quote}
\small
We need to be careful---the model does not exist for negative values of the model parameter $m^2_\nu$. (You would not know how to write a Monte Carlo simulation of detector response for negative $m^2_\nu$.) This is not the same as a zero prior; the model simply does not exist. Now, the data statistic is something that comes out of your fit and can be negative.
That is also called $m^2_\nu$ in the presentation, but should have a different name. That is why \expertLL labelled the axes in his pedagogical introduction to the Neyman construction with things that are impossible to confuse (the temperature of the Sun was the model parameter and neutrino flux counts was the data statistic). Do not call that number---the one that can be negative---``$m^2_\nu$''; call it some other data statistic. The Feldman-Cousins paper illustrates this with non-negative model parameter $\mu$ and data statistic $x$ that can be negative\footnote{However, it is conventional to call it the ``measured'' value and have the parameter name with a hat.}.
\end{quote}

This led to a comment on frequentist inference (with some pedagogical simplifications) by \expertMB:
\begin{quote}
\small
We start with a likelihood $p(y | \theta)$ which informs how the parameters,
$\theta \in \Theta$, affect possible measurements, $y \in Y$. In inference, we
want to go the other way and use explicit measurements to determine which
parameters are most consistent with those measurements. In frequentist
statistics this is done with an estimator, which is ideally a function of the
data into the parameter space, $\hat{\theta}$ : $Y \rightarrow \Theta$. The
analysis would then use the likelihood to engineer an estimator that
identifies ``true'' parameters ``on average''. This is how estimators are
usually presented, but there is no reason why estimators have to map into only
the parameter space. For example, we could define $\hat{\theta}$ : $Y
\rightarrow \Omega$, where $\Theta \subset \Omega$. Even though the estimator
does not always identify a valid point in $\Theta$, it can still identify one
often enough to work ``on average''. In the above example $\Omega$ was the
real numbers  $\mathbb{R}$, but $\Theta$ was just the positive real numbers,
$\mathbb{R}^{+}$.
\end{quote}
In other words, frequentist criteria are degenerate: there are many estimators that satisfy the criteria that are intuitive and well-behaved, and many that are not.  It was agreed that building good frequentist analyses for non-trivial problems is rarely simple.

\subsection{\label{sec:SolarNeutrinos}Solar Neutrinos}
Analysis methods used at current and recent solar neutrino experiments were described.
Because of the very low-energy nature of the events, where discrimination from backgrounds is critical, many of the statistical methods  pertained to triggering, reconstruction and event discrimination, as well as to analysis methods~\citePSnu{Smy}\footnote{The results included a frequentist answer to the question: ``what is the temperature of the Sun?''.}.
These statistical techniques included: an algorithm to searching for the interaction vertex of low-light events (a few MeV) in Super-K~\cite{Fukuda:2002uc} called BONSAI (for Branch Optimization Navigating Successive Annealing Interactions);
Hough Transforms; the separation of solar neutrinos from background by taking into account the effects of multiple scattering;
and BOREXINO's use of Gatti parameters per energy bin for $\alpha$-$\beta$ discrimination, in conjunction with a Boosted Decision Tree.

For its oscillation analysis, Super-K uses a ``mixed likelihood $\chi^2$'', which employs a binned $\chi^2$ to study the spectrum, but a likelihood function for the time dependence.
This is for reasons of computational efficiency.
The ``combined'' analysis from SNO of data from its three different phases (pure D$_2$O; with NCl; and with $^3$He counters), using six generic variables describing energy-dependent flavour changes was introduced, which parametrises the $\nu_e$ survival probability, day-night asymmetry and the total $^8$B solar neutrino flux. Super-K performs a similar analysis, but employs a standard day-night asymmetry value.
It is noteworthy that the joint analysis between SNO and Super-K takes full advantage of the fact that the uncertainty correlations are opposite to each other in the two experiments.

\subsection{\label{sec:AtmosphericNeutrinos}Atmospheric Neutrinos}
At IceCube~\cite{Achterberg:2006md} and Super-K, various different techniques are used for event selection and reconstruction, such as likelihoods, decision tree and neural networks~\citePSnu{deAndre}.
In particular, a new event reconstruction algorithm at IceCube was described, where eight event parameters are fit over likelihood distributions that are ``bumpy''. %
Here again, forward folding and ``Bayesian'' unfolding are discussed for extracting the energy spectrum at IceCube (see \cref{para:unfolding} for a discussion of unfolding).

The oscillations of atmospheric neutrinos, as observed by Super-K and IceCube-DeepCore, have both been presented in recent papers by employing Wilks' theorem, converting the $\Delta \ln L$, and profiling over the nuisance parameters.
Historically $\sin^2 2\theta$ was used as one of the oscillation parameters, but by plotting against $\sin^2\theta$, the issue of having an unphysical region is avoided---but this makes the distributions more complicated, as can be seen in the Super-K contour.%

Aside from the extraction of atmospheric oscillation parameters, it was noted that these experiments are becoming sensitive to the neutrino mass hierarchy, which is discussed in \cref{sec:MassHierarchy}.

\subsection{\label{sec:DoubleBetaDecay}Double Beta Decay}

Double beta decay experiments span a broad range of technologies and statistical techniques~\citePSnu{Shimizu}.
GERDA have performed a hypothesis test on a signal of a magnitude that corresponds to the so-called ``KK claim'' from 2004 plus background ($H_1$), with the null hypothesis ($H_0$) for the background-only case. They obtain a Bayes factor $P(H_1)/P(H_0) = 0.024$ and a $p$-value $P(N^{0\nu} = 0|H_1) = 0.01$.
They also present two lower limits on the neutrinoless-double beta decay half-life of Ge: a frequentist limit with a profile likelihood fit, and a Bayesian limit with a flat prior in $1/T$. A median sensitivity is given for both methods. CUORE also presents two lower limits.

The binned maximum-likelihood methods for EXO-200~\cite{Albert:2014awa} and KamLAND-Zen~\cite{Gando:2012jr} were also given, with the latter splitting its data into two periods, to take into account to a significant reduction of backgrounds from $^\textrm{110m}$Ag after 2012. 

In general, the experimental bounds on the decay lifetimes are obtained directly, but when the data are converted into a limit on the effective mass of the neutrino $\langle m_{\beta\beta} \rangle$, additional uncertainties need to be included, including the neutrino oscillation parameters, CP phases, and the different theoretical calculations for the nuclear matrix element (NME). KamLAND-Zen includes the effect of seven different NME calculations, taking the highest and lowest values to give the result $\langle m_{\beta\beta} \rangle < 60$--$161$ meV.

\subsection{\label{sec:BayesianEventReconstruction}Bayesian Event Reconstruction}

A new event reconstruction method for Super-K, which uses a non-parametric Bayesian algorithm is being developed~\citePSnu{Calland}, where the expression ``non-parametric'' indicates the fact that the number of parameters are not fixed\footnote{Or ``the number of things you don't know is one of the things you don't know.''}. 
This is used to handle the fact that single events at Super-K can involve any number of Cherenkov rings.
A straightforward comparison of fits with different numbers of rings does not scale well, and a method in which the normalisation term of the Bayes Theorem  is compared is often computationally intractable.
Therefore, in this method, a {\itshape mixture model} is used, with each ring being a component of the mixture (Dirichlet Prior), and reversible jump Markov Chain Monte Carlo~\cite{Green95reversiblejump} is used to sample the posterior probability of the model.
This performs not only parameter inference, but also model selection.

How the algorithm penalises the addition of more and more rings was discussed, where it was clarified that a Poisson function with a mean of one was used as the prior on the number of rings, and this makes use of what is often dubbed the ``Occam's Razor'' effect in Bayesian statistics.

In the description of the hierarchical model, it was stated that uninformative priors are used for most of the ring parameters such as position, direction and time etc., which was questioned by \expertBC.
It was stated that Gaussians with large sigmas are used as the priors, which are called ``vague'' priors.
Whether these are truly uninformative, or imply that there can be more scattering, was discussed, with the speaker stating that vertex activity is not penalised. (See \cref{para:bayesianpriorchoice} for further discussion of the choice of prior.)

\subsection{\label{sec:PeVNeutrinosAtIceCube}PeV Neutrinos at IceCube}
In addition to the neutrino oscillation measurements described above, IceCube's original design goals include neutrino astronomy, and the search for ultra-high energy (UHE) neutrinos~\citePSnu{Lu}, where the high interaction cross sections mean that signal events are characterised by down-going events  with characteristic shapes depending on the neutrino flavour.
Results in 2012 were based on simple cut-and-count analyses, while currently ongoing analyses make use of a two-dimensional (in energy and zenith angle) Poisson binned-likelihood ratio method. A comparison of the two methods was made, given some possible scenarios for the types of events that could be detected.
Hypothesis testing on astrophysical-like and GZK-like models, and the shape of the high-energy tail were also discussed, as well as a future study on whether there is a cut-off in the spectrum in the 1--10~PeV region.

Issues raised during the presentation included the likelihood ratio method, the use of Boosted Decision Trees for event classification, and blind and non-blind analyses.

Regarding the two-dimensional Poisson binned-likelihood ratio search for signal (GZK, astrophysical) neutrinos, the methods described by S.~Algeri~\citePSnu{Algeri} (see~\cite{Algeri:2015zpa, Algeri:2016gtj}) are directly applicable (\cref{sec:MassHierarchyHypothesisTesting}).

At this workshop, this was the only occasion that blind analyses were explicitly mentioned;  some of the discussion was on unblinding small amounts, say 10\%, of the data at a time, as long as it is led by good physics judgement. There was agreement that the point of a blind analysis is not that one never changes anything beyond a certain point in the analysis, but that if changes are made at a time when it is not impossible for bias to be introduced, they must be fully detailed in the description of the analysis.

\subsection{\label{sec:CosmologicalConstraints}Cosmological Constraints on Neutrino Properties}

The properties of neutrinos---principally the sum of their masses---can affect the cosmological measurements that have been advancing significantly in recent years; and so statistical methods can be used on the data to inform us about these properties~\citePSnu{Villaescusa}.

Data for the CMB power spectrum are represented as Gaussian distributions.
Various methods are used to estimate the covariances, including the production of ``mocks'' (referred to as ``toy'' experiments in particle physics).
Questions include: whether we can continue to describe the likelihood as a multivariate Gaussian; and whether we are able to produce covariance matrices that allow sub-percent level cosmological measurements to be made.

Comments included the following from \expertMB:
\begin{quote}
 \small
  You do not want to plug in a covariance estimator into the likelihood as if it were exact---that ignores uncertainty. Rather treat the covariance as unknown parameters and just fit them along with the rest of your Markov Chain. Sellentin and Heavens~\cite{Sellentin:2015waz} use a Wishart prior so that they can analytically marginalize out the covariance, but the Wishart prior has suboptimal properties and the resulting Student-$t$ distribution is extremely hard to fit with a Markov Chain anyway. We recommend the LKJ prior~\cite{Lewandowski20091989} which is has nicer properties and yields a better-behaved posterior.
  \end{quote}
  and from \expertSA:
\begin{quote}
 \small
If your covariance matrix is sparse (i.e., if you expect many of the
off-diagonal entries to be equal or close to zero) you could try some
covariance estimation methods which are strictly designed for sparse matrices.
Between those, you may want to check shrinkage estimators such as
Ledoit-Wolf~\cite{LEDOIT2004365}, thresholding methods such as adaptive thresholding and graphical
models such as the Graphical Lasso~\cite{doi:10.1093/biostatistics/kxm045}.
\end{quote}
\expertBC asked what the priors are when biases are marginalised over, with the response being that it is unclear. 
Since there are about 30 nuisance parameters, evaluating the effect of priors is computationally expensive.

  \subsection{\label{sec:GlobalFits}Global Fits of Oscillation Parameters}

Global fits of experimental data to neutrino oscillation parameters were presented from two different viewpoints---one fitting within the three-generation paradigm, and one explicitly looking for the involvement of sterile neutrinos.

\paragraph{Three-neutrino fits}For the three-neutrino case~\citePSnu{Bergstrom}, a standard likelihood and $\chi^2$-based analysis was discussed first, which produces confidence intervals and hypothesis tests. Here asymptotic distributions are assumed, but this is typically not tested, and is known to be inaccurate in some cases. 
Bayesian alternatives were then given, where data are used to update prior odds through the Bayes factor, to be used to compare models. Jeffreys' scale for the interpretation of the strength of the evidence for a model was considered.

The parameter space for three-neutrinos was discussed, and the Haar measure~\cite{Kass19961343} introduced as the unique uniform prior in the compact neutrino mass matrix space.
The specific issues raised by the mass hierarchy (see \cref{sec:MassHierarchy}), $\delta_\textrm{CP}$, specific tests for CP-violation, and $\theta_{23}$  were presented, as well as the possibility of employing Bayesian methods to evaluate the expected utility of future experiments. 
Discussion focused on the effects of $\delta$ ``wrapping around'', the differences between measuring $\delta$ as opposed to the existence of CP-violation, and the applicability of Jeffreys' scale in physics.

Other aspects of the choice of parameters for oscillations studies (empirical motivation as opposed to mathematical convenience) were also presented~\citePSnu{Litchfield}, as was the use of epicycles for the visualisation of three-flavour neutrino oscillations~\citePSnu{Xu}.

\paragraph{Sterile-neutrino fits}For the sterile neutrino fit described here~\citePSnu{Collin}, a likelihood function is computed, using 14 different oscillation results from 12 experiments (including some combined results)---a Markov Chain Monte Carlo is used, as it is not possible to scan over the entire space. Specifically, an Affine invariant parallel tempering MCMC is used.

To interpret the global results, a parameter goodness-of-fit test (PG test) is used, to compare compatibility of datasets with neutrino oscillation models---with a severe tension being observed.
Toy experiment studies indicate that experiments with significant backgrounds cause the PG test to fail.
An example of a Bayesian evaluation was shown, which may overcome this difficulty, and advice was sought. 
Some issues that need to be considered are that the global fit assumes no correlations across experiments in their nuisance parameters, which are not included in the MCMC, and the fact that some experimental results include oscillation model assumptions (for example, SciBooNE and MiniBooNE assume no $\nu_e$ appearance). 
It was also noted that some results, such as from solar experiments and Super-K, which would disfavour sterile neutrinos, are not included in the fit.

\expertBC commented that there are strong connections between the PG test and the Kalman filter. The underlying principle is the same.
\expertLL noted that there are similarities to the handling of parton distribution functions.
\expertMB stated that there is no Bayesian equivalent to the PG test, but Hierarchical modelling is one appropriate approach.
 
  \subsection{\label{sec:MassHierarchy}Neutrino Mass Hierarchy}

In this section we focus on the discussion of the statistical aspects of the
neutrino mass hierarchy\footnote{We note here that many in the community,
including some of the organisers of \phystatnu, strongly prefer the expression
``mass ordering''---but in this document we employ the word ``hierarchy'', as
this was the term that appeared to dominate in the talks and discussions. This
should not be considered as an endorsement of the term---and we look forward to
the data soon rendering the concept obsolete.}, which were a theme throughout
the three days of the workshop. Theoretical and experimental approaches,
and especially sensitivity studies in the latter case, were presented.

\subsubsection{\label{sec:MassHierarchyHypothesisTesting}Hypothesis Testing}
The mass hierarchy determination is a binary classification test between two 
non-nested 
hypotheses\footnote{The word ``disjoint'' was used in some of the discussions, %
but it was agreed by the statisticians that ``non-nested'' is the appropriate expression.}~\citePSnu{Ciuffoli}.  The standard test statistic or classification function used to
test the hypotheses is:
\begin{equation}
\Delta\chi^{2} = \chi^2_\IH-\chi^{2}_\NH,
\end{equation}
where $\chi^2_\IH$ and $\chi^{2}_\NH$ are the $\chi^2$ values of the data under
the two hypotheses respectively. This test statistic does
not follow a one-degree-of-freedom $\chi^2$ distribution, %
but in the absence of degeneracies, it is well approximated by a Gaussian distribution
of mean $\mu=\pm\overline{\Delta\chi^2}$ and width $\sigma=2\sqrt{\overline{\Delta\chi^2}}$. %

Three approaches for assessing the strength of evidence regarding the mass hierarchy have been taken in the neutrino literature:
\begin{enumerate}
  \item{Test each hypothesis and compare the likelihoods.  This method may reject or accept both hypotheses~\cite{Qian:2012zn, Blennow:2013oma}.}
  \item{Test the hypothesis that both the normal and inverted hierarchies are equally effective hypotheses.  This approach give about half the significance of other approaches~\cite{Capozzi:2013psa}.}
  \item{Use Bayesian model selection with the Bayes factor.  This approach assumes that one of the hierarchies is true~\cite{Qian:2012zn, Ciuffoli:2012bs, Blennow:2013kga}.}
\end{enumerate}
In the third approach, the formula for the sensitivity of the median experiment ({\it i.e.} where the $\Delta\chi^2$ is equal to the expected $\overline{\Delta\chi^2}$), under the
Gaussian approximation for the $\Delta\chi^{2}$ distribution was derived and presented:
\begin{equation}
s_{v}=\sqrt{2}\,\mathrm{erf}^{-1}\left(\frac{1-e^{-\overline{\Delta\chi^{2}}/2}}{1+e^{-\overline{\Delta\chi^{2}}/2}}\right).
\end{equation}
The formula for the probability to achieve a given significance was also derived.

The effect of non-linear systematic uncertainties in reactor experiments was also studied.  Nuisance
parameters with quadratic, exponential and worst-case energy reconstruction dependence
were studied, and it was found that the exponential and worst-case scenarios can significantly
impact the $\Delta\chi^{2}$.

Bayesian model selection and classical approaches have also been considered in the
context of three neutrino mixing fits~\citePSnu{Bergstrom}.  The
Bayesian approach seems to be appropriate since all parameters are common to both hypotheses and global
data have no preference (equal prior odds).  In the classical approach, the test statistic
distributions can be calculated for both hypotheses, but it was shown that the distributions
depend on the assumed true values for other oscillation parameters, particularly $\delta$.  
Hence, $p$-value calculations will depend on the assumptions of the true values of the 
oscillation parameters.  

The idea of a nested model with a parameter $\lambda$ that is used to make a weighted sum
of the predictions from each hypothesis was also presented~\cite{Algeri:2015zpa}:
\begin{equation}
n = \lambda n_\IH(\Theta)+(1-\lambda)n_\NH(\Theta).
\end{equation}
In this approach the fitted interval of $\lambda$ is used to include or exclude one or both
of the hypotheses. 

\subsubsection{\label{sec:MassHierarchyKM3NETORCA}KM3NeT/ORCA}
ORCA is a 5.7 Mt water Cherenkov detector consisting of an array of photo-detectors in the
deep Mediterranean Sea talk by~\citePSnu{Bruijn}.  It will 
detect atmospheric neutrinos with the primary goal of mass hierarchy determination.  A
study of the mass hierarchy sensitivity of ORCA was presented, in which the 
classical approach was taken, where the $\chi^2_\IH$ and $\chi^{2}_\NH$ were evaluated
on a set of pseudo-experiments and the usual $\Delta\chi^{2}$ or log-likelihood-ratio
test statistic was used.  The median significance for the wrong hierarchy exclusion was presented as:
\begin{equation}
s = \frac{\mu_\NH-\mu_\IH}{\sigma_\IH}.
\end{equation}
Here, $\mu_\NH$ and $\mu_\IH$ are the medians of the test statistic distributions for the 
normal (correct) and inverted (wrong) hierarchies respectively and $\sigma_\IH$ is the
width of the inverted (wrong) hierarchy test statistic distribution.  

In ORCA, the best-fit value of $\theta_{23}$ depends strongly on the true hierarchy.  To
account for this in the sensitivity studies, for each true $\theta_{23}$
and hierarchy, the most likely wrong hierarchy and $\theta_{23}$ combination is determined.
This dependence is parametrized and these most likely wrong hierarchy sets are used in the
significance determination.  

\subsubsection{\label{sec:MassHierarchyJUNO}JUNO} The JUNO experiment is a medium-baseline reactor neutrino experiment that aims to determine the mass hierarchy by observing the oscillation patterns of $\nu_e$ disappearance through terms depending on $\Delta m^{2}_{31}$ and $\Delta m^{2}_{32}$~\citePSnu{Li}.  It was 
shown that the sensitivity of the JUNO experiment can be improved by more precise measurements of $\Delta m^{2}$ in muon neutrinos.  

A classical treatment of the mass hierarchy sensitivity was presented, where the Type-I ($\alpha$) %
and Type-II ($\beta$) error rates are used.  Using $\alpha$ at a fixed $\beta=0.5$ gives 
the median sensitivity.  It 
was shown that this metric closely matches the significance estimated from $\sqrt{\overline{\Delta\chi^{2}}}$.  Two methods for showing the range of possible experimental
significances where presented.  In one case, the $\sqrt{\overline{\Delta\chi^{2}}}$ is shown to describe
in more detail the possible outcomes of the experiment.  In the other approach, the power 
$(1-\beta)$ is presented for fixed $\alpha$ as a function of $\overline{\Delta\chi^{2}}$.

\subsubsection{\label{sec:MassHierarchyDiscussion}Discussion }
There was no consensus from the experts to select a single best approach for reporting  sensitivities and significances for mass hierarchy determination.  Three types of approaches received endorsements from the experts.  The classical approach of producing the expected test statistics distributions for each hypothesis using pseudo-experiments was used for the Higgs spin parity measurements at the LHC and was suggested by \expertBC, who helped devise this method.  
The main challenge to this approach in neutrino experiments is that the test statistic distributions themselves can be strongly dependent on the true values assumed for other oscillation parameter values.  The Bayesian model selection approach seems suitable for mass hierarchy determination since the choice of prior probabilities for the two hierarchies (50/50) may be relatively uncontroversial and was endorsed by \expertDvD and strongly favoured by \expertDK.  \expertMB raised some concerns with the Bayesian model selection approach: the model probabilities are difficult to calculate, and they have poor predictive performance; in other words, they are vulnerable to overfitting.  \expertMB recommended a third approach of building a bigger model that encompasses the mass hierarchy hypotheses and determining the hierarchy through parameter estimation.  For example, he suggested fitting the mass values for all three mass states, although this exact method may not be possible in oscillation experiments in practice since they are not sensitive to the absolute mass scale.  A preference for fitting, as opposed to model selection, also motivates the method presented by SA~\citePSnu{Algeri}, hence this approach has implicit endorsement from \expertDvD and \expertSA.

  \section{\label{sec:StatisticalTopics}Other Statistical Topics}
Several statistical topics attracted attention, either by being touched-on independently in the physics talks, or by prompting significant discussion, including in the Panel Discussion.
\subsection{Bayesian versus frequentist approaches}
The traditional wisdom that HEP is more naturally frequentist in outlook\footnote{Although this can be something that differs between parameter estimation and hypothesis testing.} was not much in evidence among those present (the cohort not necessarily being representative). Indeed, the only presentation directly addressing the division~\citePSnu{Biller} advocated switching entirely to a Bayesian approach.  The main thrusts of this argument were: 
\begin{itemize}
 \item that (mis-)interpretation of frequentist confidence intervals as Bayesian statements is almost inevitable, \emph{especially} in cases where the two differ substantially, as this is exactly where the frequentist picture is least intuitive.
 \item that repeated experiments are something of a luxury in HEP and, as such, the frequentist concept of being correct over many trials is somewhat academic. More commonly the question is how to make best use of a single result, even if it is suspected to be atypical.
\end{itemize}
The latter item touches on ideas about conditioning and the concept of a ``recognisable subset''. This was not discussed much during the workshop but was referenced in the summary talks.  Physicists who have not renounced frequentism and all its works should perhaps investigate~\cite{reid1995} or~\cite{Cousins:2011-upper-limits} and references therein. %

The general impression was that most of those present (both statisticians and physicists) were broadly agnostic or pragmatic~\cite{kass2006}, as summarised by \expertDvD: ``Why throw away half your toolbox?''  Some trends were apparent, however.  Bayesian approaches attracted greater interest in cases where there are discrete hypotheses (e.g. determination of the mass hierarchy), particularly since the standard asymptotic frequentist methods cannot easily be applied in these cases.  The use of Bayesian methods in event classification is also uncontroversial, but also provided an interesting demonstration of issues in Bayesian model selection (see \cref{sec:BayesianEventReconstruction}).  For more conventional analyses---such as the detection of new physics signatures---frequentist approaches are still generally the first choice, with most experiments opting for full frequentist methods in cases where asymptotic formulae are not suitable.

Although the increasing acceptance of Bayesian methods was regarded positively, a few notes of caution were sounded by the statisticians, and physicists adopting Bayesian analyses should probably familiarise themselves with known pitfalls.  Bayes factors~\cite{doi:10.1080/01621459.1995.10476572} received quite some attention as a means of hypothesis discrimination, but should not be regarded as a panacea: (e.g. from \expertSA) ``[Among statisticians] $p$-values and Bayes factors are fairly equally criticised''---unless it is for small values of $p$ and large values of the Bayes factor.
Specifically, Bayes factors are highly sensitive to the choice of priors, even when that is not the case for the posterior, are difficult to calculate, and any interpretation is not a statistical statement (\expertMB), and therefore different Bayes factors should be reported for different priors (\expertSA).
The Jeffreys-Lindley Paradox (see~\cite{Cousins:2013hry} for an introduction) was also mentioned several times, as were issues with priors when models have different-sized parameter spaces.

Selection of Bayesian priors is a perennial topic. It is a truism that there is no right answer (even ``objective'' Bayesians can come to different conclusions), and although some rules of thumb for HEP were proposed~\citePSnu{Biller}, some criticisms from the experts were raised regarding the need for coverage and other issues that have been studied in depth in the literature on this topic\label{para:bayesianpriorchoice}. The argument by Kass is important in the discussion of coverage in a Bayesian context~\cite{kass2006}. A peculiar aspect of neutrino oscillation physics was highlighted---given the large mixing in the neutrino sector it seems reasonable to many that priors for mixing should reflect the $SU(3)$ symmetry~\citePSnu{Bergstrom}.  This results in a particular choice of variables for which the prior is uniform, specifically  $\pi(\mathrm{s}^2_{12},\,\mathrm{s}^2_{23},\,\mathrm{c}^4_{13},\,\delta) = \frac{1}{2\pi}$.  Although this isn't a neutral assumption (it is closely related to the phenomenological concept of neutrino mass anarchy), theoretical or empirical motivations for prior selection are largely absent from current experimental discussion.  The \emph{de facto} practice of ``flat in the variable we are plotting'' is rather more murky when the parameters are based on a particular choice of matrix decomposition. %
Reference was made to the existing literature on subjective and objective Bayesian analysis methods to guide the choice of prior\footnote{A ``flat'' prior is not included in these.}~\cite{IRONY1997159, BergerAndPericchi, Kass19961343, berger2006, goldstein2006}.

The need to understand the sensitivity of any result to the choice of priors was reiterated at several points, but the statisticians noticeably worry less about the familiar-to-physicists example of limits in one dimension and more about the problems associated with priors in higher dimensions, which are much more subtle and cannot be ``solved'' by looking at the frequentist properties of the result. 
The question of how to compare models of different complexity (i.e. with different numbers of parameters requiring priors) was presented~\citePSnu{Ikeda} and discussed in depth.  This is non-trivial, and multiple prescriptions (Akaike Information Criteria, Bayesian Information Criteria etc.) exist, as introduced by \expertSI.
\expertSA explained how these allow one to quantify the fact that more parameters will almost always produce a better fit, and \expertMB described his own work on this topic~\cite{2015arXiv150602273B}. This involves not assuming that the likelihood is correct; a statement that was queried by \expertDK---the response to which was that models need to be validated to demonstrate that they are ``close enough''. Physicists encountering these issues are urged to consult the statistics literature such as those given here (or a statistician)\label{para:modelswithdifferingcomplexities}.

\expertMB notes that prior choice is just one assumption in a statistical model, no worse than those in the choice of likelihood.  Criticisms of prior sensitivity are equally relevant to likelihood sensitivity, and indeed the entire model needs to be critiqued to build a robust analysis.  This is one of the benefits of predictive critiques, such as posterior predictive checks and the information criteria mentioned above.

Finally, although it is by no means the only frequentist approach, the widespread use of the Feldman-Cousins method in HEP means the two terms are often treated as synonymous. Quizzed on the method in particular there was consensus among the statisticians that it was a good in principle---SA: ``It has nice features, the only issue is computation''; \expertMB: ``Feldman-Cousins uses a ratio test which is basis of modern frequentist stats---nothing wrong with it. The question is: when is it practical? In one dimension it is very good, %
but it does not work in a thousand dimensions''. \expertBC responded: ``you do not need to go to one-thousand---it is not great for three dimensions!''.

\subsection{Asymptotic limits and five-sigma}

An issue that has attracted renewed discussion across HEP  %
is the use of $5\sigma$ as the discovery criterion.  Such a high threshold is almost universally regarded as excessive by statisticians, but is still regarded as a useful precaution within the community. \expertEW summarised: ``Because we want to publish, there is a serious bias towards inappropriately reducing systematics so we insist on these thresholds to compensate''.  It was also suggested that high thresholds are a sociological side-effect of results having to filter up through collaborations before being published.  The statistician's view, stated during an early discussion prompted by \expertLL, was that ``Two $3.5\sigma$ results are better than one $5\sigma$ result''; \expertDvD went further his summary talk~\citePSnu{VanDyk} and suggested ``One \emph{calibrated} $3.5\sigma$ result is better than an uncalibrated $5\sigma$''. %

It was emphasised by \expertMB (and others) that the problem with $5\sigma$ is not so much that the required probability (1 in 2.7 million) is excessive, but that evaluating distributions to this level of precision is computationally infeasible: in order to claim an effect has a $z$-value (``significance'') of just over $5\sigma$ under a hypothesis $\mathcal{H}$ it is necessary to know the distribution of outcomes of $\mathcal{H}$ down to a precision of $\mathcal{O}(10^{-7})$.  In practice, this is usually obtainable only with the use of approximate or asymptotic methods, by far the most common being by application of Wilks' theorem.  It is rarely investigated whether the statistic used for testing has actually reached asymptotic behaviour---indeed it is often clear that it has not, for example if two-dimensional intervals drawn at fixed $\Delta(\ln\mathcal{L})$ are not elliptical. 

The computational difficulty in mapping the joint space of parameters and the data are what give rise to complaints that ``Feldman-Cousins takes too long''.  It was noted that this problem is also true of \emph{any} correct frequentist method if asymptotic formulae cannot be relied upon, Feldman-Cousins is just the go-to method in a common situation where asymptotics fail.  Furthermore---and much less appreciated---this is not a problem of frequentist methods: \expertMB pointed out that the applicability (or not) of asymptotic limits is unrelated to the Bayesian/frequentist divide, and that Bayesian inference is also incredibly fast when you can assume that the asymptotic limit is valid.  Simply mapping out the likelihood function in high-dimensions is computationally difficult, and this is the starting point for both Bayesians and frequentists.  \expertMB also noted that ``there is a large community in statistics, including myself and my colleagues, who do not use asymptotics due to these [validity] issues.''%

There does not appear to be generally-applicable way out of this predicament (apart from renouncing $5\sigma$).  However, a few methods were described that allow some relief, if applied carefully to the problem at hand.  For the testing of discrete, non-nested hypotheses, it was shown~\citePSnu{Algeri} that if the hypotheses can be embedded in a more general model, the problem can be recast in a way analogous to the calculation of trial factors used to correct for the ``Look Elsewhere Effect''.  This can allow a more robust approach to approximating significance calculation~\cite{Gross2010}.  The more general point made was that there are other asymptotic methods besides Wilks' theorem and, depending on the problem, they may work better.
The issues around \emph{efficiently} sampling high-dimensional distributions were also discussed by \expertMB~\citePSnu{Betancourt1} in the form of a pedagogical outline of various deterministic and stochastic methods (the recommendation is for scalable Markov Chain Monte Carlo methods, in particular Hamiltonian Monte Carlo).

\subsection{Profiling and marginalisation} \label{subsec:overallprofilingmarginalisation}
These methods of nuisance-parameter elimination, as seen previously (\cref{para:LBLmarginalisationprojection} and~\cref{para:reactorneutrinoprofiling}) can give very different responses depending on the distributions, although the difference is less important if they are near the Gaussian limit.
Marginalisation `makes sense' on a mathematical level, but philosophical issues exist for frequentists since it often amounts to formulating a prior for the parameter to be eliminated.  This is not a problem if that is a detector parameter that is not interesting to report, but is more subtle if one is suppressing another physics parameter (e.g., to summarise a two-dimensional contour as two one-dimensional uncertainties).  Profile likelihoods are in some sense a more natural approach for frequentists as they do not ascribe a weight to any particular value of a nuisance parameter, however they are not true likelihoods, and this occasionally causes problems.  The statisticians present agreed with \expertDK about never profiling if you are a Bayesian. They also noted that statisticians tend to frame the problem by calculating the convergence rates.

\subsection{Limited experimental acceptances\label{para:limitedacceptances}}
The question of how to present results that prefer regions in parameter space where the experiment has no sensitivity appeared at several points in the discussions.  This often overlaps with the Bayesian vs frequentist debate, because these cases tend to highlight the difference between the approaches, but the problem exists independently of that discussion.  In the introductory statistics overview~\citePSnu{Lyons}, and during the discussion, the {\itshape CL}$_s$ method~\cite{ZECH1989608, 0954-3899-28-10-313} was mentioned. Instead of hypothesis testing (or equivalently, drawing intervals) at a fixed tail probability $p_0$ for the null hypothesis, the method additionally uses the null-tending tail probability $p_1$ of an alternate signal hypothesis, and sets limits based on the value of $p_0/(1-p_1)$.  This modification has previously seen use in HEP (it was invented for LEP analyses), although it is essentially unknown in the reporting of neutrino physics results.    

\subsection{Generative modelling}
The proposal by \expertMB to introduce Generative Modelling (put simply, the passing-on of information, from physics models to experimental set-ups to data distributions, using likelihood distributions~\citePSnu{Betancourt2} led to a significant amount of discussion (see also \cref{para:generativemodelling} for its use in cross-section measurements).
It was emphasised that this does not have to be implemented from beginning to end to be useful; any point along the flow of information can be replaced by the use of likelihoods.
\expertDK raised the concern that the number of dimensions that the likelihoods would involve make it impossible to imagine it working in practice. \expertMB responded that one only needs to first define the input and outputs that are required, and to approximate. 
There was no real opposition to the idea in principle, but practical questions of how to describe a function in so many dimensions, and to what extent this really would introduce robustness against drastic model changes could not be answered during this workshop.

\subsection{Machine learning\label{para:machinelearning}}
One question that arose during discussions was how to communicate results that rely on machine learning techniques~\citePSnu{Gillies}.  It was recognised that dealing with systematic uncertainties can be particularly tricky, especially if the machine learning is used at a high level of the analysis.  This is another topic that ought to be expanded on in future workshops.
In particular, \expertMB stated that in a proper Bayesian analysis all proper statistical answers are given by expectations with respect to the posterior, including summary statistics and marginalization, but machine learning methods do not. Consequently it is very hard to incorporate machine learning into your analysis unless they can be interpreted as generative models~\cite{bishop2006pattern}.  When a method is not generative then it has to be modelled as part of the data reduction itself, such as with a detector trigger or any other selection procedure.

\subsection{Terminology} 
In the Panel Discussion, some time was spent on how to improve communication between physicists and statisticians, in particular, the use of terminology, which can be confusing.\footnote{A example of possible confusion came up during the workshop itself: the term `projection'. Some take it to be synonymous with `profiling', others use it to refer to a simpler approach---resulting in some discussion at crossed purposes.} Should physicists all be using terms such as ``prior predictive'', ``posterior predictive'', ``plug-in method'', ``supremum method'' (all terms that arise in hypothesis testing)?

It was stated by the statisticians that too often, physicists make up names themselves for pre-existing concepts, which could be avoided if they spoke to statisticians more often. Rather than the ``statistics committees'' that large experiments have, a more general interface between the fields was preferred. 
Physics should feature in statistics conferences, and vice versa\footnote{For the first time, at Neutrino 2016, which was held a month after this workshop, a statistics talk was given by a statistician (\expertDvD).}.
\expertEW suggested that as there are many theorists on the DUNE Collaboration in particular, this could be extended to statisticians.
\expertSA added that an alternative would be if physicists published everything in the form of likelihoods---this would certainly reduce confusion.
But in general, there was a consensus that more communication between neutrino physicists and statisticians would be valuable, continuing the discussions held at this workshop.

\section{\label{sec:Summary}Summary of Discussions}

Here we list the main outcomes and open questions from the discussion at \phystatnuipmu.
\begin{itemize}
  \item Prefer marginalisation over profiling or projection when eliminating nuisance parameters (\cref{para:LBLmarginalisationprojection}) in Bayesian analyses---but complications relating to implied priors may arise when marginalising away any parameter of interest in a frequentist context.
  \item Refine use of prior- and posterior-predictive methods for studying systematic effects (\cref{para:priorpredictive}).
  \item Discuss the use of Bayes factors (\cref{para:bayesfactors}) and the Jeffreys scale in neutrino physics, where the consensus at this workshop was to use these with caution. Consider also the Jeffreys-Lindley Paradox.
  \item Physicists are strongly urged to study the statistical literature on subjective and objective Bayesian priors (\cref{para:bayesianhierarchicalmodelling}). Neutrino physics in particular includes areas where phenomenological concepts can have an influence on the meaning of priors.
  \item The comparison of models in high dimensions, and choosing between models of differing complexity---such that different numbers of parameters require priors---is an area where the statistical literature on information criteria should be consulted further (\cref{para:modelswithdifferingcomplexities}).
  \item Be very careful with ``unfolding'' of detector effects---for example, when presenting cross-section data---and also any subsequent refolding in of detector effects (\cref{para:unfolding}).
  \item Consider use of dedicated software tools to allow the release of measurement information in more modern ways (\cref{para:datareleasetools}).
  \item Strive to move towards publishing cross section measurement information in ways that are statistically sound and more useful for those who will use them, such as the ones presented here (\cref{para:crosssectionpublishedmeasurement})?
  \item Consider making more use of Generative Modelling methods (\cref{para:generativemodelling}) more when publishing cross sections and other results.
  \item Consider whether Bayesian Hierarchical Modelling could be helpful when making use of external cross-section data and multiple cross-section models (\cref{para:bayesianhierarchicalmodelling}), compared to current methods.
  \item Be prepared for experiments to be faced by the problem of favouring regions in parameter space in which an experiment has no sensitivity (\cref{para:limitedacceptances})? 
  \item Should the {\itshape CL}$_s$ method be introduced in neutrino physics? Other methods that are given in the statistical literature received significant support (\cref{para:limitedacceptances}).
  \item The presentation of results that rely on using machine learning techniques at a high level of analysis requires further discussion \cref{para:machinelearning}). There are important statistical issues, such as compatibility with Bayesian methods, which need to be addressed.
  \item Topics for further in-depth discussion include: the presentation of results in a statistically useful way for future physicists; formal methods of integrating machine learning methods with the statistical analysis techniques that are in use; and the streamlining of the language used by physicists to describe statistical concepts.
\end{itemize}
\paragraph{Acknowledgements}We thank the Kavli IPMU for its generous support and the provision of its facilities which allowed the workshop to be a success. The authors of this paper are also supported by the Science and Technology Facilities Council, UK and the Swiss National Science Foundation and SERI, Switzerland.

\bibliographyPSnu{PhyStat-nu-ConferenceBibliography}
\renewcommand{\refname}{References}
\bibliographystyle{hunsrt}
\bibliography{NeutrinoBibliography}

\appendix

\section{\label{sec:Participants}List of Participants}

\small
Sara Algeri (Imperial College London),
Josh Amey (Imperial College London),
Mohammad Sajjad Athar (Aligarh Muslim University),
Joao Pedro Athayde Marcondes De Andre (Michigan State University),
Christopher Barry (University of Liverpool),
Johannes Bergstr\"om (Universitat de Barcelona),
Michael Betancourt (University of Warwick),
Christophe Bronner (Kavli IPMU),
Ronald Bruijn (University of Amsterdam/Nikhef),
Richard Calland (Kavli IPMU),
Son Cao (Kyoto University),
Daniel Cherdack (Colorado State University),
Wonqook Choi (Korea Institute of Science and Technology information),
Georgios Christodoulou (University of Liverpool),
Emilio Ciuffoli (IMP, CAS),
Gabriel Collin (Massachusetts Institute of Technology),
Robert Cousins (UCLA),
Andrew Cudd (Michigan State University),
Stephen Dolan (Oxford University),
Patrick Dunne (Imperial College London),
Arturo Fiorentini (York University),
Megan Friend (KEK),
Monojit Ghosh (Tokyo Metropolitan University),
Ewen Gillies (Imperial College London),
Francesco Gizzarelli (CEA Saclay),
Le\"ila Haegel (University of Geneva),
Mark Hartz (Kavli IPMU (WPI), University of Tokyo/TRIUMF),
Yoshinari Hayato (Kamioka Observatory, ICRR, University of Tokyo),
Katsuki Hiraide (ICRR, University of Tokyo),
Shiro Ikeda (Institute of Statistical Mathematics),
Asher Kaboth (Royal Holloway University of London),
Michiru Kaneda (Tokyo Institute of Technology),
Dean Karlen (University of Victoria and TRIUMF),
Teppei Katori (Queen Mary University of London),
Marat Khabibullin (INR RAS),
Marco Kleesiek (Karlsruhe Institute of Technology),
Youngju Ko (Chung-Ang University),
Eunhyang Kwon (Seoul National University),
Mathieu Lamoureux (CEA, IRFU),
Pierre Lasorak (Queen Mary University of London),
Phillip Litchfield (Imperial College London),
Yufeng Li (Institute of High Energy Physics),
Tianmeng Lou (University of Tokyo),
Livia Ludhova (IKP FZJ),
Lu Lu (Chiba University),
Louis Lyons (Imperial College London),
Paul Martins (Queen Mary University of London),
Wing Yan Ma (Imperial College London),
Mark Mccarthy (York University),
Mikio Morii (Institute of Statistical Mathematics),
Shigetaka Moriyama (ICRR, University of Tokyo),
Jacob Morrison (Michigan State University),
Keigo Nakamura (Kyoto University),
Christine Nielsen (University of British Columbia),
Yasuhiro Nishimura (ICRR, University of Tokyo),
Yoomin Oh (Institute for Basic Science),
Kimihiro Okumura (ICRR, University of Tokyo),
Jose Palomino (Stony Brook University),
Seokhee Park (Yonsei University),
Luke Pickering (Imperial College London),
Elder Pinzon (York University),
Ciro Riccio (University of Naples ),
Gabriel Santucci (Stony Brook University),
Hiroyuki Sekiya (ICRR,  University of Tokyo),
Hyunkwan Seo (Seoul National University),
Kyungmin Seo (Institute for Basic Science, Korea),
Sunny Seo (Seoul National University),
Davide Sgalaberna (University of Geneva),
Raj Shah (Oxford/RAL),
Ralitsa Sharankova (Tokyo Institute of Technology),
Itaru Shimizu (Research Center for Neutrino Science, Tohoku University),
ChangDong Shin (Chnnam National University),
Michael Smy (University of California, Irvine),
Amit Kumar Srivastava (D.A.V. College Kanpur, India),
Patrick Stowell (University of Sheffield),
Yoshi Uchida (Imperial College London),
Zoya Vallari (Stony Brook University),
David Van Dyk (Imperial College London),
Cristovao Vilela (Stony Brook University),
Francisco Villaescusa-Navarro (INAF-Trieste),
Tomislav Vladisavljevic (University of Oxford),
Yue Wang (Stony Brook University),
Morgan Wascko (Imperial College London),
Callum Wilkinson (University of Bern),
Elizabeth Worcester (Brookhaven National Laboratory),
Clarence Wret (Imperial College London),
Benda Xu (IPMU, University of Tokyo),
InSung Yeo (Chnnam National University),
Tomoyo Yoshida (Tokyo Institute of Technology),
Mitchell Yu (York University)

\section{\label{sec:PhyStatNuSeries}The \phystatnu Series of Workshops}

This document summarises the first \phystatnu, held at the Kavli IPMU near Tokyo between the 30th of May and 2nd of June 2016. The website for this workshop is at \url{http://conference.ipmu.jp/PhyStat-nu}.

Following this, a sister workshop was held at Fermilab on the 19th till the 21st of September 2016. Records of the workshop are maintained at \url{https://indico.fnal.gov/event/11906/}.

The participants at the original \phystatnu strongly favoured the continuation of the workshop series, and indicated that about every three years would be appropriate for its frequency. A \phystatnu is scheduled for the 23rd to the 26th of January 2019 at CERN.

\section{\label{sec:RecommendedReading}Recommended Introductory Reading}
Below is a selected list of introductory reading that has been particularly recommended by the experts who participated in the workshop.
\begin{itemize}
  \item ``Data Reduction and Error Analysis for the Physical Sciences'', by Philip R.~Bevington and D.~Keith Robinson
    \url{http://highered.mheducation.com/sites/0072472278/}
    A quick read for an undergraduate-level review.
  \item ``Statistical Data Analysis'', by Glen Cowan
    \url{http://www.pp.rhul.ac.uk/~cowan/sda/}
    A solid foundation for High Energy Physics analysis methods.
  \item ``Information Theory, Inference, and Learning Algorithms'', by David Mackay
    \url{http://www.inference.phy.cam.ac.uk/itila/book.html}
    A broad-ranging book, as indicated by its title, but for statistical learning purposes, Chapters 1, 2, 3, and Section IV (except perhaps for Chapters 25 and 26) are very useful.
  \item ``Pattern Recognition and Machine Learning'', by Chris Bishop
    \url{http://www.springer.com/us/book/9780387310732}
    For machine learning algorithms and Bayesian inference, with strong focus on algorithms, including pseudocode examples. 
  \item ``Statistical Rethinking: A Bayesian Course with Examples in R and Stan''
    \url{http://xcelab.net/rm/statistical-rethinking/}
    A brand new book that introduces Bayesian inference with little mathematics, instead focusing on concepts, modelling techniques, and their implementation in software. The models discussed are pretty simple, but many different techniques are reviewed, and this could be a good primer for ``Bayesian Data Analysis''.
  \item Bayesian Reading List (\expertBC): References \cite{Koopman1943-KOOJHT, BergerAndPericchi, Kass19961343, berger2006, goldstein2006, IRONY1997159, doi:10.1080/01621459.1989.10478756} constitute a selection of reading that is recommended for physicists. The last of these in particular focuses on model selection.
\item ``Bayesian Data Analysis'', by Andrew Gelman et al.
 \url{http://www.stat.columbia.edu/~gelman/book/}
A major reference work for many statisticians.
The nomenclature can take some time penetrate, but this is one of the few books that introduces many of the most powerful modelling techniques in use today. Anyone who digests the entire book will be strongly rewarded.
\item ``Bayes Factors'', by Robert Kass and Adrian Raftery''~\cite{doi:10.1080/01621459.1995.10476572},
At 24 pages, a brief and comprehensive introduction to Bayes Factors.
\item Recommendations and notes from the CDF Statistics Committee: The CDF Statistics Committee, from which many of the ideas in the original ``collider'' PhyStat workshops arose, maintains a list of reading (\url{https://www-cdf.fnal.gov/physics/statistics/statistics_recommendations.html}) and technical notes (\url{https://www-cdf.fnal.gov/physics/statistics/statistics_cdfnotes.html}).

\end{itemize}

\vspace{5mm}
\end{document}